\documentclass[twocolumn,showpacs,preprintnumbers,amsmath,amssymb]{revtex4}
\usepackage{docs}

\usepackage{graphicx}
\usepackage{dcolumn}
\usepackage{dsfont}
\usepackage{bm}

\begin{document}
\title{General class of vacuum Brans-Dicke wormholes}

\author{Francisco S. N. Lobo}
\email{flobo@cii.fc.ul.pt} \affiliation{Centro de Astronomia e
Astrof\'{\i}sica da Universidade de Lisboa, Campo Grande, Ed. C8
1749-016 Lisboa, Portugal}

\author{Miguel A. Oliveira}
\email{miguel@cosmo.fis.fc.ul.pt}\affiliation{Centro de Astronomia
e Astrof\'{\i}sica da Universidade de Lisboa, Campo Grande, Ed. C8
1749-016 Lisboa, Portugal}

\date{\today}

\begin{abstract}

Recently, traversable wormhole geometries were constructed in the
context of $f(R)$ gravity. The latter is equivalent to a
Brans-Dicke theory with a coupling parameter $\omega=0$, which is
apparently excluded from the narrow interval, $-3/2<\omega<-4/3$,
extensively considered in the literature of static wormhole
solutions in vacuum Brans-Dicke theory. However, this latter
interval is only valid for a specific choice of an integration
constant of the field equations derived on the basis of a
post-Newtonian weak field approximation, and there is no reason
for it to hold in the presence of compact objects with strong
gravitational fields. In this context, we construct a general
class of vacuum Brans-Dicke wormholes that include the value of
$\omega=0$.

\end{abstract}

\pacs{04.20.Jb, 04.50.Kd}

\maketitle

{\it Introduction}: Wormholes are hypothetical tunnels in
spacetime and in classical general relativity are supported by
exotic matter, which involves a stress energy tensor that violates
the null energy condition (NEC) \cite{Morris:1988cz}. Several
candidates have been proposed in the literature, such as solutions
in Einstein-Gauss-Bonnet theory \cite{EGB}; wormholes on the brane
\cite{braneWH}; solutions in Brans-Dicke theory
\cite{Agnese:1995kd,Anchordoqui:1996jh,Nandi:1997en,Nandietal,Nandietal2},
which will be further explored in this brief report; wormhole
solutions in semiclassical gravity \cite{Garattini:2007ff}; exact
wormhole solutions using conformal symmetries
\cite{Boehmer:2007rm}; solutions supported by equations of state
responsible for the cosmic acceleration \cite{phantomWH}; and NEC
respecting geometries were further explored in conformal Weyl
gravity \cite{Lobo:2008zu}; the possibility of distinguishing
wormhole geometries by using astrophysical observations of the
emission spectra from accretion disks was also explored
\cite{Harko:2008vy}, etc (see Refs.
\cite{Lemos:2003jb,Lobo:2007zb} for more details and
\cite{Lobo:2007zb} for a recent review).

Recently, traversable wormhole geometries in the context of $f(R)$
modified theories of gravity were also constructed
\cite{Lobo:2009ip}. The matter threading the wormhole was imposed
to satisfy the energy conditions, so that it is the effective
stress-energy tensor containing higher order curvature derivatives
that is responsible for the NEC violation. Thus, the higher order
curvature terms, interpreted as a gravitational fluid, sustain
these nonstandard wormhole geometries, fundamentally different
from their counterparts in general relativity. Furthermore, we
note that $f(R)$ modified theories of gravity are equivalent to a
Brans-Dicke theory with a coupling parameter $\omega=0$, and a
specific potential related to the function $f(R)$ and its
derivative. However, the value $\omega=0$ is apparently excluded
from the interval, $-3/2<\omega<-4/3$, of the coupling parameter,
extensively considered in the literature of static wormhole
solutions in vacuum Brans-Dicke theory.

In Brans-Dicke theory, analytical wormhole solutions were
constructed \cite{Agnese:1995kd,Anchordoqui:1996jh,Nandi:1997en}.
It was shown that static wormhole solutions in vacuum Brans-Dicke
theory only exist in a narrow interval of the coupling parameter
\cite{Nandi:1997en}, namely, $-3/2<\omega<-4/3$. However, this
result is only valid for vacuum solutions and for a specific
choice of an integration constant of the field equations given by
$C(w)=-1/(\omega+2)$. The latter relationship was derived on the
basis of a post-Newtonian weak field approximation, and it is
important to emphasize that there is no reason for it to hold in
the presence of compact objects with strong gravitational fields.

In this context, we construct a general class of vacuum
Brans-Dicke wormholes that include the value of $\omega=0$, and
thus constructing a consistent bridge with the wormhole solutions
in $f(R)$ gravity found in \cite{Lobo:2009ip}. Furthermore, we
present the general condition for the existence of Brans-Dicke
wormhole geometries based on the NEC violation, and show that the
presence of effective negative energy densities is a generic
feature of these vacuum solutions.

\medskip


{\it General class of Brans wormholes}: The matter-free action in
Brans-Dicke theory is given by
\begin{equation}
S=\frac{1}{2}\int d^{4}x(-g)^{\frac{1}{2}}\left[ \varphi
\mathbf{R} -\varphi ^{-1}\omega (\varphi )g^{\mu \nu }\varphi
_{,\mu }\varphi _{,\nu } \right] \,,
\end{equation}
where $\mathbf{R}$ is the curvature scalar, $\omega $ is a
constant dimensionless coupling parameter, and $\varphi$ is the
Brans-Dicke scalar. We adopt the convention $8\pi G=c=1$
throughout this work.

The above action provides the following field equations:
\begin{eqnarray}
\mathbf{G}_{\mu \nu }&=&-\frac{\omega }{ \varphi ^{2}}\left(
\varphi _{,\mu }\varphi _{,\nu }-\frac{1}{2}g_{\mu \nu }\varphi
_{,\sigma }\varphi ^{,\sigma }\right)
   \nonumber  \\
&&  -\frac{1}{\varphi }\left( \varphi _{;\mu }\varphi _{;\nu
}-g_{\mu \nu }\square ^{2}\varphi \right) \,,  \\
\square ^{2}\varphi &=&0\,,
\end{eqnarray}
where $\mathbf{G}_{\mu \nu }$ is the Einstein tensor and $\square
^{2}\equiv \varphi ^{;\rho }{}_{;\rho }$.

It is useful to work in isotropic coordinates, with the metric
given by
\begin{equation}
ds ^{2}=-e^{2\alpha (r)}dt^{2}+e^{2\beta (r)}dr^{2}+e^{2\nu
(r)}r^{2}(d\theta ^{2}+\sin ^{2}\theta d\psi ^{2}).
\label{isometric}
\end{equation}
Throughout this work, we consider the Brans class I solution,
which corresponds to setting the gauge $\beta -\nu =0$. Thus, the
field equations yield the following solutions
\begin{eqnarray}
e^{\alpha (r)}&=&e^{\alpha _{0}}\left(
\frac{1-B/r}{1+B/r}\right)^{\frac{1}{ \lambda }} \label{feqs1}
\,,\\
e^{\beta (r)}&=&e^{\beta _{0}}\left(1+B/r\right) ^{2}\left(
\frac{1-B/r}{1+B/r }\right)^{\frac{\lambda -C-1}{\lambda }}\,,
\label{feqs2}
\\
\varphi (r)&=&\varphi _{0}\left(\frac{1-B/r}{1+B/r}\right)
^{\frac{C}{\lambda }}, \label{feqs3}
\\
\lambda ^{2}&\equiv & (C+1)^{2}-C\left( 1-\frac{\omega
C}{2}\right)>0\,, \label{feqs4}
\end{eqnarray}
where $\alpha _{0}$, $\beta _{0}$, $B$, $C$, and $\varphi _{0}$
are constants. Note that the asymptotic flatness condition imposes
that $\alpha _{0}=$ $\beta _{0}=0$, as can be readily verified
from Eqs. (\ref{feqs1}) and (\ref{feqs2}).

In order to analyze traversable wormholes in vacuum Brans-Dicke
theory, it is convenient to express the spacetime metric in the
original Morris-Thorne canonical form \cite{Morris:1988cz}:
\begin{equation}
ds^{2}=-e^{2\Phi (R)}dt^{2}+\frac{dR^{2}}{1-b(R)/R}+R^{2}(d\theta
^{2}+\sin ^{2}\theta d\psi ^{2})
\end{equation}
where $\Phi (R)$ and $b(R)$ are the redshift and shape functions,
respectively. To be a wormhole solution, several properties are
imposed \cite{Morris:1988cz}, namely, the throat is located at
$R=R_0$ and $b(R_0)=R_0$. A flaring out condition of the throat is
imposed, i.e., $[b(R)-Rb'(R)]/b^2(R)>0$, which reduces to
$b'(R_0)<1$ at the throat, where the prime here denotes a
derivative with respect to $R$. The condition $1-b(R)/R\geq 0$ is
also imposed. To be traversable, one must demand the absence of
event horizons, so that $\Phi(R)$ must be finite everywhere.

Confronting the Morris-Thorne metric with the isotropic metric
(\ref{isometric}), the radial coordinate $r\rightarrow R$ is
redefined as
\begin{equation}
R=re^{\beta _{0}}\left(1+B/r\right)^{2}\left(
\frac{1-B/r}{1+B/r}\right)^{\Omega }\,,\qquad\Omega
=1-\frac{C+1}{\lambda }\,, \label{defineR}
\end{equation}
so that $\Phi(R)$ and $b(R)$ are given by
\begin{equation}
\Phi (R)=\alpha _{0}+\frac{1}{\lambda }\left\{ \ln \left[
1-\frac{B}{r(R)} \right]-\ln \left[1+\frac{B}{r(R)}\right]
\right\} , \label{redshift}
\end{equation}
\begin{equation}
b(R)=R\left\{ 1-\left[ \frac{\lambda
[r^{2}(R)+B^{2}]-2r(R)B(C+1)}{\lambda
[r^{2}(R)+B^{2}]}\right]^{2}\right\} \,,\label{shape}
\end{equation}
respectively. The wormhole throat condition $b(R_{0})=R_{0}$
imposes the minimum allowed $r$-coordinate radii $r_{0}^{\pm }$
given by
\begin{equation}
r_{0}^{\pm }=\alpha ^{\pm }B\,, \qquad  \alpha ^{\pm }=(1-\Omega
)\pm \sqrt{\Omega (\Omega -2)}\,. \label{throatr0}
\end{equation}
The values $R_{0}^{\pm }$ can be obtained from Eq. (\ref{defineR})
using Eq. (\ref{throatr0}). Note that $R\rightarrow \infty $ as
$r\rightarrow \infty $, so that $b(R)/R\rightarrow 0$ as
$R\rightarrow \infty $. The condition $ b(R)/R\leq 1$ is also
verified for all $R\geq $ $R_{0}^{\pm }$. The redshift function
$\Phi (R)$ has a singularity at $r=r_{S}=B$, so that the minimum
allowed values of $r_{0}^{\pm }$ must necessarily exceed
$r_{S}=B$. It can also be verified from Eq. (\ref{defineR}) that $
r_{0}^{\pm }\geq B$ which implies $R_{0}^{\pm }\geq 0$.


The energy density and the radial pressure of the wormhole
material are given by \cite{Nandi:1997en}
\begin{eqnarray}
\rho&=&-\frac{4B^{2}r^{4}Z^{2}[(C+1)^{2}-\lambda^{2}]}{\lambda^{2}(r^{2}
-B^{2})^{4}}\,,\label{rhoR}  \\
p_{r}&=&-\frac{4Br^{3}Z^{2}}{\lambda^{2}(r^{2}-B^{2})^{4}}
[\lambda C(r^{2} +B^{2})
    \nonumber   \\
&&-Br(C^{2}-1+\lambda^{2})] \label{prR}\,,
\end{eqnarray}
respectively, where $Z$ is defined as
\begin{equation}
Z\equiv\left(  \frac{r-B}{r+B}\right)^{(C+1)/\lambda}.
\end{equation}

Adding Eqs. (\ref{rhoR}) and (\ref{prR}), one arrives at
\begin{equation}
\rho+p_{r}=-\frac{4Br^{3}Z^{2}}
{\lambda^{2}(r^{2}-B^{2})^{4}}[\lambda
C(r^{2}+B^{2})+2Br(C+1-\lambda^{2})]\,, \label{NECviolation}
\end{equation}
which will be analyzed in the NEC violation below.

In \cite{Nandi:1997en}, the authors considered negative energy
densities, which consequently violates the weak energy condition
(WEC). Now, Eq. (\ref{rhoR}) imposes the following condition:
\begin{equation}
\left[C(\omega )+1\right]^2>\lambda^2 (\omega )\,,
   \label{Cwlambda}
\end{equation}
which can be rephrased as
\begin{equation}
C(\omega )\left[ 1-\frac{\omega C(\omega )}{2}\right] >0\,,
   \label{Cwlambda2}
\end{equation}
by taking into account Eq. (\ref{feqs4}). Note that the function
$C(\omega)$ is still unspecified.

However, it is important to emphasize that negative energy
densities are not a necessary condition in wormhole physics. The
fundamental ingredient is the violation of the NEC, $\rho+p_r<0$,
which is imposed by the flaring out condition
\cite{Morris:1988cz}. To find the general restriction for
$\rho+p_{r}<0$ at the throat $r_0$, amounts to analyzing the
factor in square brackets in Eq. (\ref{NECviolation}), namely, the
condition $\lambda C(r_0^{2}+B^{2})+2Br_0(C+1-\lambda^{2})>0$.
Using Eqs. (\ref{feqs4}) and (\ref{throatr0}), the latter
condition is expressed as:
\begin{eqnarray}
&&(-1)^{s+t+1}\left[ (-1)^s(C+1)+(-1)^t\sqrt{C\bigg(1-\frac{\omega
C}{2}\bigg)}\right]\times
  \nonumber   \\
&&\times\frac{C \left(1-\omega
C/2\right)}{\sqrt{(4+2\omega)C^2+4(C+1)}}>0 \,,\label{gen-cond}
\end{eqnarray}
where $s,t=0,1$. Note that a necessary condition imposed by the
term in the square root, in square brackets, is precisely
condition (\ref{Cwlambda2}). Thus, a necessary condition for
vacuum Brans-Dicke wormholes is the existence of negative
effective energy densities. However, we emphasize that it is
condition (\ref{gen-cond}), i.e., the violation of the NEC at the
throat, that generic vacuum Brans-Dicke wormholes should obey.


A specific choice of $C(\omega)$ considered extensively in the
literature, is the Agnese-La Camera function \cite{Agnese:1995kd}
given by
\begin{equation}
C(\omega )=-\frac{1}{\omega +2}\,.\label{Cw}
\end{equation}
Using this function, it was shown that static wormhole solutions
in vacuum Brans-Dicke theory only exist in a narrow interval of
the coupling parameter \cite{Nandi:1997en}, namely,
$-3/2<\omega<-4/3$. However, we point out that this result is only
valid for vacuum solutions and for the specific choice of
$C(\omega)$ considered by Agnese and  La Camera
\cite{Agnese:1995kd}. As mentioned in the Introduction,
relationship (\ref{Cw}) was derived on the basis of a
post-Newtonian weak field approximation, and it is important to
emphasize that there is no reason for it to hold in the presence
of compact objects with strong gravitational fields. The choice
given by (\ref{Cw}) is a tentative example and reflects how
crucially the wormhole range for $\omega$ depends on the form of
$C(\omega)$. Evidently, different forms for $C(\omega)$ different
from Eq. (\ref{Cw}) would lead to different intervals for
$\omega$.

Note that in \cite{Nandietal}, the negative values of the coupling
parameter $\omega$ were extended to arbitrary positive values of
omega, i.e., $\omega<\infty$, in the context of two-way
traversable wormhole Brans solutions (we refer the reader to Ref
\cite{Nandietal} for specific details). An interesting example was
provided in Ref. \cite{Matsuda:1972uk}, in the context of
gravitational collapse in the Brans-Dicke theory, where the choice
$C(\omega)\sim-\omega^{-1/2}$ was analyzed. More specifically, the
authors in \cite{Nandi:1997en} considered
$C(\omega)=-q\omega^{-1/2}$, with $q<0$ so that $C(\omega)>0$.
Thus, the constraint (\ref{Cwlambda2}) is satisfied only if
$\omega> 4/q^2$. However, we will be interested in solutions which
include the value $\omega=0$, in order to find an equivalence with
the $f(R)$ solutions found in \cite{Lobo:2009ip}. The specific
choices we consider below possess the following limits, $C(\omega
)\rightarrow 0$, $ \lambda (\omega )\rightarrow 1$ as $\omega
\rightarrow \infty $, in order to recover the Schwarzschild
exterior metric in standard coordinates.

Another issue that needs to be mentioned is that the
above-mentioned interval imposed on $\omega$ was also obtained by
considering negative energy densities. In principle, the violation
of the WEC combined with an adequate choice of $C(\omega)$ could
provide a different viability and less restrictive interval
(including the value $\omega=0$) from the case of
$-3/2<\omega<-4/3$ considered in \cite{Nandi:1997en}. In this
context, we consider below different forms of $C(\omega)$ that
allow the value $\omega=0$ in the permitted range. Thus, to
satisfy the constraint (\ref{Cwlambda2}), both factors $C(\omega)$
and $[1-\omega C(\omega )/2]$ should both be positive, or both
negative.

\begin{figure}[h]
\centering
  \includegraphics[width=2.3in]{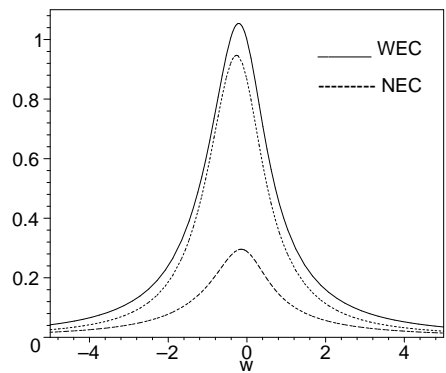}
  \caption{Plot of the energy conditions for
  $C(\omega)=(\omega^2+a^2)^{-1}$ for $a=1$.
  In particular, the WEC expressed by condition (\ref{Cwlambda2})
  is given by the solid line; and the NEC, expressed by the
  condition (\ref{gen-cond}), is given by the dashed curves. For the
latter, only the cases of $(s,t)=(0,1)$ and $(s,t)=(1,1)$ of
condition (\ref{gen-cond}) are allowed; and are depicted by the
small and large peaks, respectively.}
 \label{Fig:C-hat}
\end{figure}

Consider the following specific choice
\begin{equation}
C(\omega )=\frac{1}{\omega^2+a^2}\,,\label{Cw2}
\end{equation}
where $a$ is a real constant. The requirement that
$\lambda^{2}>0$, i.e., Eq. (\ref{feqs4}), is satisfied. The
function $C(\omega)$ is positive for all real $\omega$, and the
second term, in square brackets, of Eq. (\ref{Cwlambda2}), is
positive everywhere for $a^2>1/16$. Therefore, for this case,
condition (\ref{Cwlambda2}) is satisfied for all $\omega$. For
$a^2<1/16$, $[1-\omega C(\omega)/2]$ has two real roots, namely,
$\omega^0_\pm=(1\pm\sqrt{1-16a})/4$; the lesser value is positive
and thus both the second term and condition (\ref{Cwlambda2}) will
be positive at $\omega=0$. Thus, if $a^2<1/16$, the condition
(\ref{Cwlambda2}) is satisfied for $\omega\in \mathds{R}-
[\omega^0_-;\omega^0_+]$. Figure~\ref{Fig:C-hat} depicts condition
(\ref{Cwlambda2}) (depicted as a solid curve), i.e., negative
energy densities, and condition (\ref{gen-cond}) (depicted as the
dashed curves), i.e., the violation of the NEC, for $a=1$. For the
latter, only the cases of $(s,t)=(0,1)$ and $(s,t)=(1,1)$ of
condition (\ref{gen-cond}) are allowed; and are depicted in
Fig.~\ref{Fig:C-hat} by the small and large peaks, respectively.

In the limiting case, $C(\omega )\rightarrow 0$, $ \lambda (\omega
)\rightarrow 1$ as $\omega \rightarrow \infty $, one simply
recovers the Schwarzschild exterior metric in standard
coordinates. This can be verified from Eqs. (\ref{redshift}) and
(\ref{shape}), which impose $b(R)=2M$ and $b'|_{r_0}=0$. However,
in this limit, the inequality (\ref{gen-cond}) is violated, and
there are no traversable wormholes.

Consider a second specific choice given by
\begin{equation}
C(\omega)=A\exp\left(-\frac{\omega^2}{2}\right)\,.
\label{Cw3}
\end{equation}
The requirement that $\lambda^{2}>0$, i.e., Eq. (\ref{feqs4}), is
also satisfied. This function, for $A>0$, is positive for all
$\omega$. Therefore, in order to satisfy condition
(\ref{Cwlambda2}), the restriction $(1-\omega C(\omega)/2)>0$ is
imposed. We verify that if $0<A<2\exp(1/2)$, then $(1-\omega
C(\omega)/2)>0$ for all $\omega$, so that conditions
(\ref{Cwlambda2}) and (\ref{gen-cond}) are both satisfied. If
$A>2\exp(1/2)$, then the second term $(1-\omega C(\omega)/2)$ will
have two real positive roots, i.e., $\omega_{0,1}>0$. For this
choice of $A$, we have the following range of allowed $\omega$:
$\mathds{R}-]\omega_0,\omega_1[$. Moreover, since $\omega_0>0$,
the value $\omega=0$ will always be in the set of allowed values.

\begin{figure}[h]
\centering
  \includegraphics[width=2.3in]{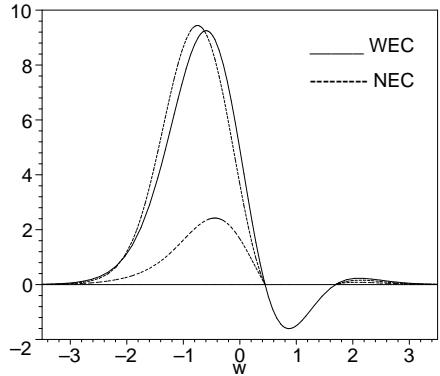}
  \caption{Plot of the energy conditions for
  $C(\omega)=A\exp(-\omega^2/2)$,
  with $A=3\exp(1/2)$.
  In particular, the WEC expressed by condition (\ref{Cwlambda2})
  is given by the solid line; and the NEC, expressed by the
  condition (\ref{gen-cond}), is given by the dashed curve.
  For the latter, only the cases of $(s,t)=(0,1)$ and $(s,t)=(1,1)$
  of condition (\ref{gen-cond}) are allowed; and are depicted by
  the smaller and larger peaks, respectively.}
 \label{Fig:2}
\end{figure}

Figure~\ref{Fig:2} depicts condition (\ref{Cwlambda2}) (depicted
as a solid curve), i.e., negative energy densities, and condition
(\ref{gen-cond}) (depicted as dashed curves), i.e., the violation
of the NEC for $A=3\exp(1/2)$. For the latter, only the cases of
$(s,t)=(0,1)$ and $(s,t)=(1,1)$ of condition (\ref{gen-cond}) are
allowed; and are depicted in Fig.~\ref{Fig:2} by the smaller and
larger peaks, respectively.


{\it Conclusion}: Recently, in the context of $f(R)$ modified
theories of gravity, traversable wormhole geometries were
constructed. As $f(R)$ gravity is equivalent to a Brans-Dicke
theory with a coupling parameter $\omega=0$, one may be tempted to
find these solutions inconsistent with the permitted interval,
$-3/2<\omega<-4/3$, extensively considered in the literature of
static wormhole solutions in vacuum Brans-Dicke theory. Thus the
choice provided by Eq. (\ref{Cw}), in addition to the WEC and NEC
violation, reflects how crucially the range of $\omega$ depends on
the form of $C(\omega)$, and we have shown that adequate choices
of $C(\omega)$ provide different viability regions and less
restrictive intervals, that include $\omega=0$. In this context,
we have constructed a more general class of vacuum Brans-Dicke
wormholes that include the value of $\omega=0$, proving the
consistency of the solutions constructed in $f(R)$ gravity.
Furthermore, we deduced the general condition for the existence of
vacuum Brans-Dicke wormhole geometries, and have shown that the
presence of effective negative energy densities is a generic
feature of these vacuum solutions. It will also be interesting to
generalize this analysis in Brans-Dicke theory in the presence of
matter. Work along these lines is presently underway.



{\it Acknowledgements}: MO acknowledges financial support from a
grant attributed by Centro de Astronomia e Astrof\'{\i}sica da
Universidade de Lisboa (CAAUL), and financed by Funda\c{c}\~ao
para a Ci\^{e}ncia e Tecnologia (FCT).




\begin{thebibliography}{99}

\bibitem{Morris:1988cz}
  M.~S.~Morris and K.~S.~Thorne,
  Am.\ J.\ Phys.\  {\bf 56}, 395 (1988).

\bibitem{EGB}
B. Bhawal and S. Kar, Phys. Rev. D {\bf 46}, 2464-2468 (1992);
%
G. Dotti, J. Oliva, and R. Troncoso, Phys. Rev. D {\bf 75}, 024002
(2007).

\bibitem{braneWH}
L. A. Anchordoqui and S. E. P Bergliaffa, Phys. Rev. D {\bf 62},
067502 (2000);
%
K. A. Bronnikov and S.-W. Kim, Phys. Rev. D {\bf 67}, 064027
(2003);
%
M. La Camera, Phys. Lett. {\bf B573}, 27-32 (2003);
%
F.~S.~N.~Lobo,
  Phys.\ Rev.\ D {\bf 75}, 064027 (2007).

\bibitem{Agnese:1995kd}
  A.~G.~Agnese and M.~La Camera,
  Phys.\ Rev.\  D {\bf 51}, 2011 (1995).

\bibitem{Anchordoqui:1996jh}
  L.~A.~Anchordoqui, S.~E.~Perez Bergliaffa and D.~F.~Torres,
  Phys.\ Rev.\  D {\bf 55}, 5226 (1997).

\bibitem{Nandi:1997en}
K.~K.~Nandi, B.~Bhattacharjee, S.~M.~K.~Alam and J.~Evans,
  Phys.\ Rev.\  D {\bf 57}, 823 (1998).

\bibitem{Nandietal}
K.~K.~Nandi, A.~Islam and J.~Evans,
  Phys.\ Rev.\  D {\bf 55}, 2497 (1997);
%
\bibitem{Nandietal2}
A.~Bhattacharya, I.~Nigmatzyanov, R.~Izmailov and K.~K.~Nandi,
  Class.\ Quant.\ Grav.\  {\bf 26} (2009) 235017.



\bibitem{Garattini:2007ff}
S. V. Sushkov,
Phys. Lett. \textbf{A164}, 33-37 (1992);
%
A. A. Popov,
Phys. Rev. D \textbf{64}, 104005 (2001);
%
D. Hochberg, A. Popov and S. V. Sushkov,
Phys. Rev. Lett. \textbf{78}, 2050 (1997);
%
N. R. Khusnutdinov and S. V. Sushkov,
Phys. Rev. D \textbf{65}, 084028 (2002);
%
A. R. Khabibullin, N. R. Khusnutdinov and S. V. Sushkov,
Class. Quant. Grav. \textbf{23} 627-634 (2006);
%
N. R. Khusnutdinov,
Phys. Rev. D \textbf{67}, 124020 (2003);
%
R.~Garattini and F.~S.~N.~Lobo,
  Class.\ Quant.\ Grav.\  {\bf 24}, 2401 (2007);
%
R.~Garattini and F.~S.~N.~Lobo,
  Phys.\ Lett.\  B {\bf 671}, 146 (2009).

\bibitem{Boehmer:2007rm}
  C.~G.~Boehmer, T.~Harko and F.~S.~N.~Lobo,
  Phys.\ Rev.\  D {\bf 76}, 084014 (2007);
%
C.~G.~Boehmer, T.~Harko and F.~S.~N.~Lobo,
  Class.\ Quant.\ Grav.\  {\bf 25}, 075016 (2008).

\bibitem {phantomWH}
S.~Sushkov, Phys. Rev. D {\bf 71}, 043520 (2005);
%
F.~S.~N.~Lobo, Phys.\ Rev.\ D {\bf 71}, 084011 (2005);
%
  F.~S.~N.~Lobo, Phys.\ Rev.\ D {\bf 71}, 124022 (2005);
%
F.~S.~N.~Lobo, Phys.\ Rev.\ D {\bf 73}, 064028 (2006);
%
F.~S.~N.~Lobo, Phys.\ Rev.\  D {\bf 75}, 024023 (2007);
%
P. F. Gonz\'alez-D\'{\i}az, Phys. Rev. D {\bf 68}, 084016 (2003);
%
P. F. Gonz\'{a}lez-D\'{i}az, Phys. Rev. Lett. {\bf 93} 071301
(2004);
%
P. F. Gonz\'{a}lez-D\'{i}az and J. A. Jimenez-Madrid, Phys. Lett.
{\bf B596} 16-25 (2004).

\bibitem{Lobo:2008zu}
  F.~S.~N.~Lobo, Class.\ Quant.\ Grav.\  {\bf 25}, 175006 (2008).

\bibitem{Harko:2008vy}
  T.~Harko, Z.~Kovacs and F.~S.~N.~Lobo,
  Phys.\ Rev.\  D {\bf 78}, 084005 (2008);
%
T.~Harko, Z.~Kovacs and F.~S.~N.~Lobo,
  Phys.\ Rev.\  D {\bf 79}, 064001 (2009).

\bibitem{Lemos:2003jb}
  J.~P.~S.~Lemos, F.~S.~N.~Lobo and S.~Quinet de Oliveira,
  Phys.\ Rev.\  D {\bf 68}, 064004 (2003).

\bibitem{Lobo:2007zb}
  F.~S.~N.~Lobo,
  ``Exotic solutions in General Relativity: Traversable wormholes and 'warp
  drive' spacetimes,''
  arXiv:0710.4474 [gr-qc].

\bibitem{Lobo:2009ip}
  F.~S.~N.~Lobo and M.~A.~Oliveira,
  Phys.\ Rev.\  D {\bf 80}, 104012 (2009).


\bibitem{Matsuda:1972uk}
  T.~Matsuda,
  Prog.\ Theor.\ Phys.\  {\bf 47}, 738 (1972).




\end{thebibliography}
\end{document}